\documentclass[aps,prl,groupedaddress,nofootinbib,floatfix,twocolumn,showpacs]{revtex4}

\usepackage{graphicx}

\begin{document}

\title[Crystalline Undulator]{Crystalline Undulator with a Small Amplitude and a Short Period}

\author{Andriy Kostyuk}


\affiliation{65933 Frankfurt am Main, Germany}

\begin{abstract}
The crystalline undulator is a single crystal with periodically bent 
crystallographic planes. If ultrarelativistic charged particles
channel through such a crystal, they emit hard radiation of undulator
type. 
A crystalline undulator with a
bending amplitude smaller than the distance between the bent 
planes and a bending period shorter than the period of channeling 
oscillations is proposed. 
Heretofore, it was believed that such a range 
of bending parameters was unsuitable for a crystalline undulator.
This point of view is refuted. 
In fact,
the undulator with a small 
amplitude and a short period is far superior to
what was proposed previously. 
It requires much lower beam energy for 
production of photons of the same frequency.
Such an undulator allows for a larger effective number of 
undulator periods. It is predicted to emit intense undulator radiation 
in the forward direction
with a narrow spectral distribution and a lower and softer background.
The undulator effect is seen for both positron and electron beams. 
Using positrons is, however, preferable because they enable one to 
obtain higher intensity of the undulator radiation with lower background.
\end{abstract}

\pacs{61.85.+p, 02.70.Uu, 41.75.Fr,41.60.Ap}
\maketitle


Synchrotron radiation sources are important tools of modern science. Their applications
range from solid state physics, material science, and the semiconductor industry to molecular 
biology and medicine (see e.g. \cite{Willmott2011book}).
The present state-of-the-art sources can emit coherent photons
with energy of up to $\sim10$ keV. In the regime of spontaneous emission,
the photon energy can reach
hundreds of keV  \cite{Schmueser2008book}.
Moving to higher photon energies
 would be very difficult or even impossible within the present technology.
Utilizing the phenomenon of particle channeling  \cite{Lindhard1965}
may solve this problem: electrostatic fields inside a crystal are very strong and they can 
undulate trajectories of
channeling particles much more effectively than even the best superconductive magnets.
Channeling takes place if a particle enters a crystal at a small angle to major crystallographic 
planes or axes \cite{Lindhard1965}. The particle becomes confined by the (inter)planar or (inter)axial potential 
and moves along
the corresponding plane or axis performing, respectively, transverse oscillations or rotations 
around the potential
minimum. The radiation that is produced due to the transverse motion of a charged projectile
is known as the channeling 
radiation \cite{Kumakhov:1976ti}. It has much in common with the undulator radiation, but because the shape of the 
transverse potential is not parabolic, the motion is not harmonic. As the result, the spectrum
of the channeling radiation is broader than that of the undulator radiation, especially in the case of negatively charged
projectiles.

It was suggested that a crystal with periodically bent crystallographic planes can be used to 
obtain radiation  of undulator type with a narrower spectral distribution \cite{Kaplin1980,Baryshevsky1980}.
In its initial form, the idea of the crystalline undulator assumed that the projectile 
should follow the sinusoidal shape of the bent crystallographic planes performing 
channeling oscillations at the 
same time around the bent central plane of the channel. This implied that 
the undulator period had to be much larger than the period of channeling oscillations
\begin{equation}
\lambda_\mathrm{u} \gg \lambda_\mathrm{c}.
\label{large_lambdau}
\end{equation}
Later, a condition of a large amplitude $a_\mathrm{u}$
of the crystal bending was proposed  \cite{Korol:1998ga}
\begin{equation}
a_\mathrm{u} \gg d.
\label{large_a}
\end{equation}
Here $d$ is the distance between
the bent crystallographic planes that form the channel, i.e. it is
the width of the channel. In the following, the crystalline undulator satisfying conditions
(\ref{large_lambdau}) and (\ref{large_a}) is called a 
\textbf{l}arge \textbf{a}mplitude 
and \textbf{l}ong \textbf{p}eriod \textbf{c}rystalline \textbf{u}ndulator
(LALP CU).
The complete list of conditions that have to be satisfied by the parameters of LALP CU
can be found in \cite{Korol:2004ug}.  

It was realized \cite{Korol:1999im} that the dechanneling phenomenon 
imposes severe restriction on the effective number of periods $N_\mathrm{u}$ of the 
crystalline undulator. The latter should be large, $N_\mathrm{u} \gg 1$,
to ensure a narrow spectral distribution of the undulator radiation.

The dechanneling takes place due to
incoherent collisions of the channeling particle with crystal constituents \cite{Lindhard1965}.
Being randomly scattered, the projectile on the average gains transverse energy. If the 
latter exceeds the height of the interchannel potential barrier, 
the particle leaves the channel. 
Starting from this point, it does 
not follow the shape of the channel and, consequently, it does not emit the undulator radiation. 
For this reason, the effective number of undulator periods is limited by the average 
length $L_d$ at which the dechanneling\footnote{Similarly, the effective number of periods is limited 
by the attenuation of photons
in the crystal medium if the attenuation length is smaller than the dechanneling length, 
$L_\mathrm{a} < L_\mathrm{d}$. The photon attenuation in CU was discussed  
for the first time in \cite{Baryshevsky1980}. } takes place:
\begin{equation}
1 \ll N_\mathrm{u} \simeq \frac{L_d}{\lambda_\mathrm{u}},
\label{Nu_Ld}
\end{equation}
where $\lambda_\mathrm{u}$ is the bending period of the crystalline undulator.

One more condition is relevant to the present discussion. It ensures
a stable channeling of the projectile in the bent crystal 
\cite{Tsyganov1976,Baryshevsky1980}. It is convenient to write it down in the 
form\footnote{The conditions (\ref{large_lambdau}), (\ref{large_a}) and (\ref{stable_ch}) 
are tightly related. In particular, (\ref{large_lambdau}) can be derived from 
(\ref{large_a}) and (\ref{stable_ch}) \cite{Korol:2004ug}.
Alternatively, the centrifugal parameter can be expressed in the form 
$C=R_\mathrm{c}/R_{\min}$, where $R_\mathrm{c}=E/U'_{\max}$ in the critical Tsyganov
radius of the channel \cite{Tsyganov1976} and 
$R_{\min}=\lambda_\mathrm{u}^{2}/(4 \pi^{2} a_\mathrm{u})$ is the minimal curvature
radius of a sinusoid with the amplitude $a_\mathrm{u}$ and the period $\lambda_\mathrm{u}$.  
} 	
\begin{equation}
1 > C \equiv \frac{F_\mathrm{cf}}{U'_{\max}} =
4 \pi^{2} \frac{a_\mathrm{u} E}{\lambda_\mathrm{u}^{2} U'_{\max}} .
\label{stable_ch}
\end{equation}
Here $C$ is the centrifugal parameter \cite{Korol:2004ug},
$F_{\rm cf}$ is the centrifugal force acting on the 
projectile in the bent channel, $U'_{\max}$
is the maximal force that keeps the particle in the channel and
$E$ is the energy of the projectile.

It appears to be difficult to satisfy conditions (\ref{large_a})--(\ref{stable_ch}) 
simultaneously. In particular, they cannot be satisfied in the case of
electron beam of moderate energy, $E \lesssim 1$ GeV.
The LALP CU conditions can be fulfilled
for positron beams (see \cite{Korol:2004ug} and references therein)
and for high energy, $E > 10$ GeV, electrons
\cite{Tabrizi:2006yi}. 
Nevertheless, the total energy of  channeling radiation
(integrated over the frequency)
exceeds that of the undulator radiation  even in these cases 
(see e.g. figure 8 of \cite{Korol:2004ug} or figure 4 of \cite{Tabrizi:2006yi}). 
Moreover, the channeling photons are harder than the undulator ones
and, therefore, they cannot be easily screened out. 
This may cause serious problems for many potential applications.

\vspace{3mm}

Let us revisit conditions (\ref{large_a}) and (\ref{stable_ch}) and 
check if they 
are indeed necessary to construct an intense source of hard photons
with a narrow spectral distribution. 
The further analysis will combine theoretical approach with Monte Carlo
simulations using the computer
code ChaS ({\bf Cha}nneling {\bf S}imulator). The code performs 
a 3D simulation of particle trajectories and calculates the
spectral and angular distribution of the emitted radiation.
In contrast to other channeling codes that either use the continuous potential approximation
\cite{Artru:1990nz,Biryukov:1995hv,Bogdanov_Mathematica_2010,Guidi_2010,Korol:2001ir,Saitoh_code_1985,Taratin_code_1979}
or consider binary collisions of the projectile with the crystal atoms as a whole (ignoring the 
incoherent collisions with atomic electrons) 
\cite{Robinson1963,Kudrin_code_1973,Andersen:1979qs,Bak:1984fw,Smulders_code_1987}, 
the algorithm of ChaS is based on the binary collisions of the projectile with target 
electrons as well as with nuclei.
This novel 
feature is especially beneficial in the case of negatively charged projectiles, which
channel in the vicinity of the atomic nuclei, where the continuous potential approximation 
becomes less accurate and the electron density is much higher than the average one.
Results obtained with previous versions of the code 
were published in \cite{Kostyuk:2010hs,Kostyuk:2011kh} and demonstrated reasonable agreement 
with experimental data. The present version of ChaS uses the electron distribution in the crystal
that has been calculated from first principles using the computer code ABINIT 
\cite{abinit} and utilizes more efficient algorithms than the previous versions.
It takes into account all the
physics that is essential for the analysis of the channeling of electrons
and positrons with energy $E$ in the range from a few hundreds of MeV to a few GeV,
provided that the photons under interest are not very soft: $\hbar \omega \gtrsim 0.5$ MeV
and at the same time are not very hard $\hbar \omega \ll E$. The contributions of the 
phenomena that are not included in the model (e.g. a shift of photon energy due 
to recoil,  losses of the projectile 
energy due to emission of photons, quantum effects in the motion of the projectile,
influence of the crystal medium on the emission, propagation of the radiation)
are expected to be small and cannot influence the result substantially 
\cite{Kostyuk:2010hs,Uggerhoj:2005ms}.
In the following simulations, the emittance of the particle beam was neglected, i.e. the
particles were assumed to enter the crystal at zero angle to the crystallographic planes.

Let us reanalyze the reasons behind condition (\ref{large_a}). 
The channel width $d$ sets the upper bound of the amplitude of the channeling 
oscillations. The amplitude of undulator oscillations has to be much larger
than that of the channeling oscillations to make sure that the spectrum is dominated
by the undulator radiation rather than by the channeling one. This is, however, necessary
only if the frequency of the undulator radiation $\omega_\mathrm{u}$ is
smaller than the frequency of channeling oscillations $\omega_\mathrm{c}$.
Indeed, the energy radiated in the forward direction by a moving particle in the 
case of dipole approximation has the following dependence on the transverse
oscillation amplitude $a$ and the radiation frequency $\omega$:
\begin{equation}
\left.
\frac{d \mathcal{E}}{d \omega \, d \Omega} 
\right |_{\theta=0}
\sim a^2 \omega^4,
\label{omega4}
\end{equation}
where $d \Omega$ is the differential of the solid angle
and $\theta$ is the angle between the direction of the 
radiation emission and the average direction of the particle
motion.\footnote{The dependence (\ref{omega4}) becomes obvious
from Larmor's formula for the radiated power applied to
the nonrelativistic harmonic oscillations of the projectile
in the inertial frame
that moves with the average velocity of the particle
along the beam direction.
Boosting the radiation to the laboratory frame scales the 
frequency by the factor of $2 \gamma$ but it does not change
the proportionality $\sim \omega^4$.
}
One sees from (\ref{omega4}) that  
condition (\ref{large_a}) is not necessary; i.e.,
the amplitude of the undulator bending can be smaller than the 
channel width,
\begin{equation}
a_\mathrm{u} < d,
\label{small_a}
\end{equation}
provided that the frequency of the undulator radiation is substantially 
larger than that of the channeling radiation
\begin{equation}
\omega_\mathrm{u} \gg  \omega_\mathrm{c}.
\label{large_omegau}
\end{equation}
In this case, however, the
period of the crystal bending $\lambda_\mathrm{u}$ has to be much smaller
than the smallest period of channeling oscillations $\lambda_\mathrm{c}$:
\begin{equation}
\lambda_\mathrm{u} \ll \lambda_\mathrm{c}.
\label{small_lambdau}
\end{equation}
The last inequality seems to lead to violation of condition (\ref{stable_ch}).
Indeed, the length of the period of the channeling oscillations
can be estimated as 
\begin{equation}
\lambda_\mathrm{c} \simeq 2 \pi \sqrt{\frac{E}{U''(0)}}.
\label{lambdac}
\end{equation}
Here $U''(0)$ is the second derivative of the transverse
potential energy with respect to $y$ in the point of its minimum $y=0$, 
where the axis $y$ is perpendicular to
the channel boundaries.  
Combining (\ref{lambdac}) with (\ref{small_lambdau}) and 
taking into account that\footnote{Expressions (\ref{lambdac}) and (\ref{Upmax})
become exact equalities in the case of parabolic potential. For a real potential, 
the second derivative varies and the maximum value of the force is reached 
at $|y|<d$, hence (\ref{lambdac}) is an approximate equality and (\ref{Upmax})
is an inequality.} 
\begin{equation}
U'_{\max} \lesssim U''(0)  d
\label{Upmax}
\end{equation}
one obtains from (\ref{stable_ch})
\begin{equation}
1 > C \gg  \frac{a_\mathrm{u}}{d}.
\label{Cad}
\end{equation}
The crystal bending amplitude $a_\mathrm{u}$ cannot be much smaller than $d$ otherwise 
it becomes comparable to (or smaller than) the amplitude of thermal vibrations
of the crystal atoms. No undulator effect can be observed in this case.
If $a_\mathrm{u}$ is smaller but comparable to $d$, inequality (\ref{Cad}) becomes contradictory. 
One may think that this should destroy the undulator effect, but,
fortunately, it does not happen. Condition (\ref{stable_ch}) is, in fact,
irrelevant in the case of a \textbf{s}mall \textbf{a}mplitude 
(\ref{small_a}) and \textbf{s}hort \textbf{p}eriod (\ref{small_lambdau})
(SASP CU). 

\begin{figure}[tb]
\includegraphics[width=0.99\linewidth]{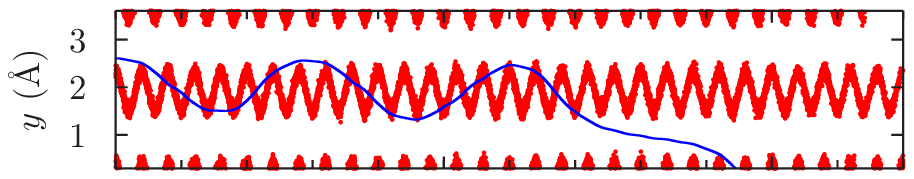}\\[2mm]
\includegraphics[width=0.99\linewidth]{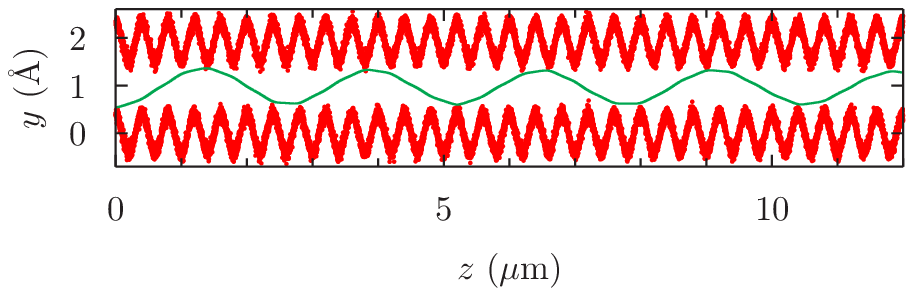}
\caption{Simulated trajectories of 
an electron (upper panel) and
a  positron (lower panel) with energy $E=855$ MeV
channeling in a crystalline undulator
with a small amplitude, $a_\mathrm{u}=0.4$ \AA, and a short period, 
$\lambda_\mathrm{u}=400$ nm. The projectile does not 
follow the shape of the bent crystallographic planes (the thick wavy lines)
instead it performs channeling oscillations with roughly the same period 
as in a straight crystal. The effect of crystal bending on the shape of trajectories
is practically unseen. \label{traj}}
\end{figure}

Simulated trajectories of an electron and a positron  channeling in 
a SASP CU are shown
in figure \ref{traj}. As is seen, the particles do not follow the
shape of the bent crystallographic planes. 
Therefore, formula (\ref{stable_ch}) is not applicable.
The channeling process is not destroyed.
Roughly,
the particle motion can be considered as if it were
governed by a continuous potential averaged
over the oscillations of the plane. In other words, it is similar
to the channeling in a straight crystal with a somewhat different
continuous potential. 

\begin{figure}[bt]
\includegraphics[width=0.99\linewidth]{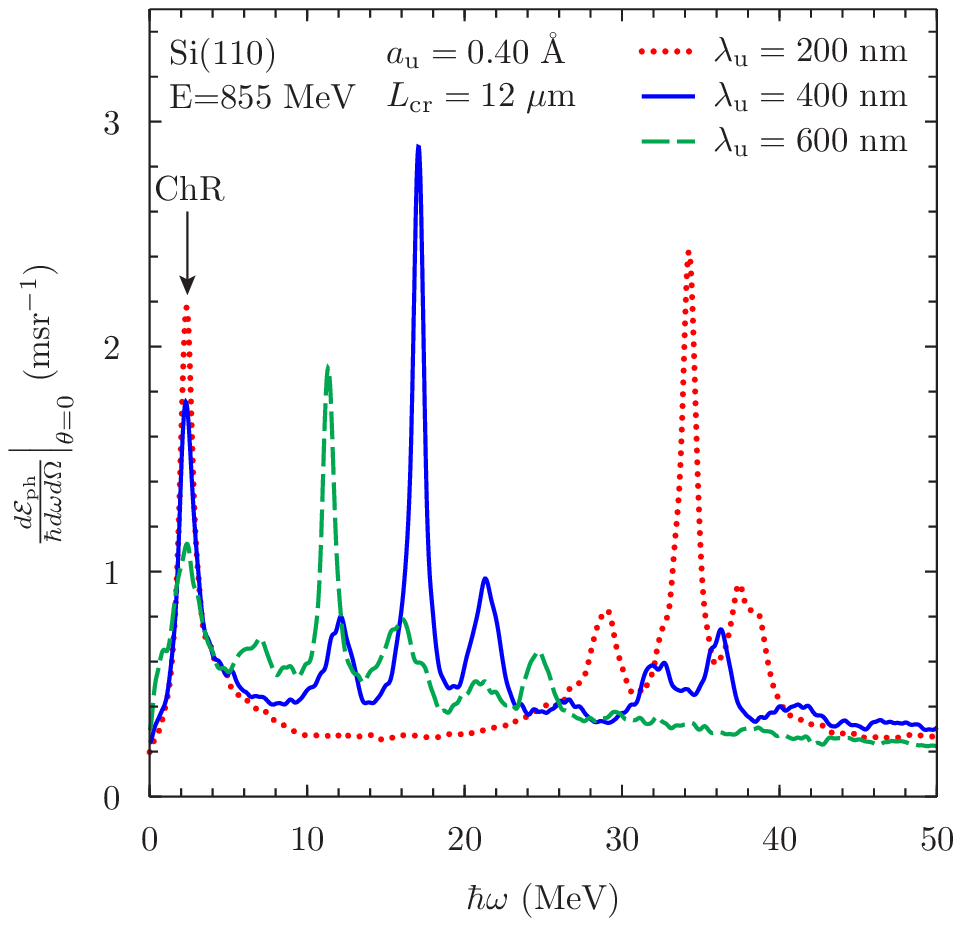}
\caption{Simulated spectra of radiation emitted by $E=855$ MeV
electrons channeling 
in a $12$ $\mu$m long
crystalline undulator with a small bending amplitude for three
short periods. The spectra are obtained by averaging over about
1000 simulated trajectories.
The undulator radiation peaks
are higher
and are centered at much larger photon energy
than the corresponding channeling radiation (ChR) peaks. 
\label{e-_spectra}}
\end{figure}

\begin{figure}[tb]
\includegraphics[width=0.99\linewidth]{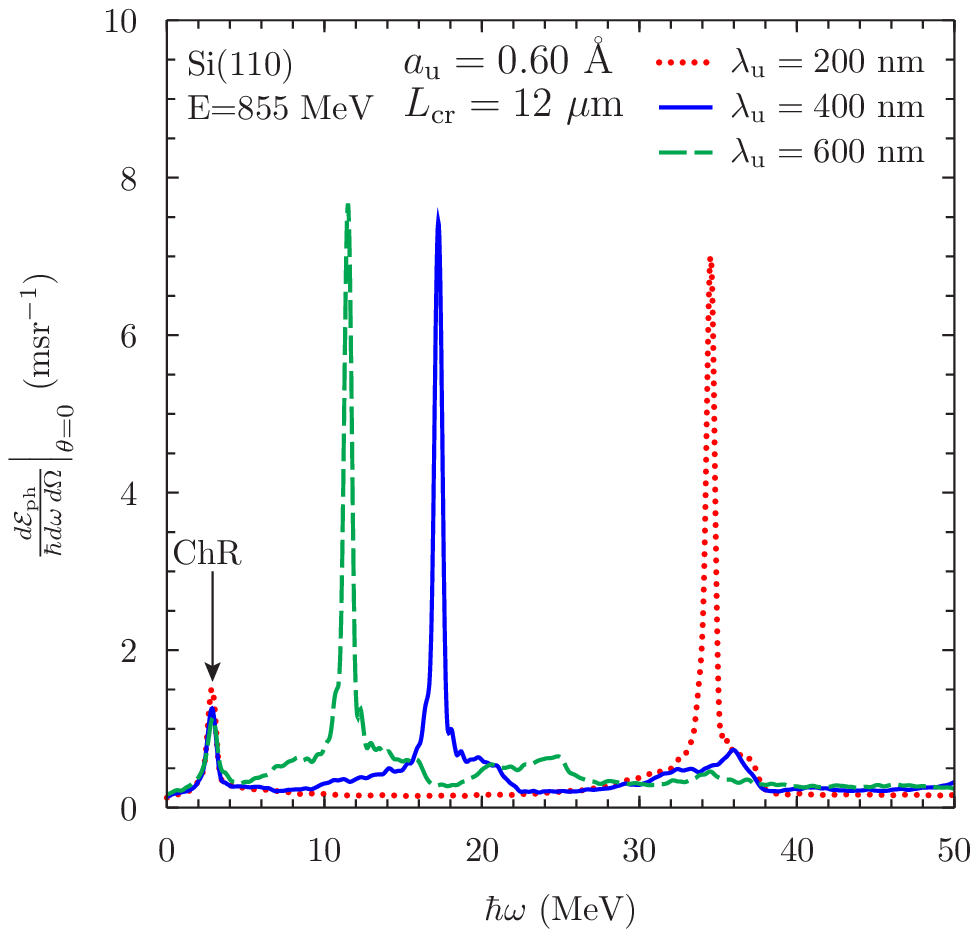}
\caption{The same as in figure \ref{e-_spectra}
but for positrons. The undulator radiation peaks
are higher than in figure \ref{e-_spectra}  by a factor 
of $\sim 3$, a better signal-to-background ratio is predicted.
\label{e+_spectra}}
\end{figure}

Nonetheless, a more careful consideration 
reveals that the particles also perform transverse oscillations 
with the period $\lambda_\mathrm{u}$. The amplitude of 
these undulator 
oscillations $a$ is much smaller than the bending amplitude 
$a_\mathrm{u}$.
Therefore, it is practically impossible to see the
modification of the trajectories due to the crystal 
bending in figure \ref{traj}.
Still, the corresponding Fourier harmonics reveal themselves
in the spectra of the emitted radiation shown in figures 
\ref{e-_spectra} and \ref{e+_spectra}.
Despite that the amplitude $a$ of the undulator 
oscillations of the projectile
is much smaller than that of the channeling oscillations,
the undulator radiation peak 
is  enhanced relative to the channeling radiation maximum
due to (\ref{omega4}) and (\ref{large_omegau}).
The undulator peaks are narrow and well separated from the 
channeling ones. The 
\textit{absolute} width of the undulator and channeling 
radiation peaks is approximately the same.
But the \textit{relative} width of undulator peaks
are much smaller due to (\ref{large_omegau}).

It is important to note that there exist technologies 
that enable one to manufacture a SASP CU. The most mature 
of them is the growing of 
Si$_{1-\chi}$Ge$_\chi$ crystals \cite{Breese:1997va}
with a periodically varying Ge content $\chi$ \cite{Mikkelsen:2000ky}.

The crystal has to be grown in the direction [001] with the 
atomic fraction $\chi$ of
germanium 
varying between $0$ and $2 \overline{\chi}$
according to the formula \cite{Kostyuk2013}
\begin{equation}
\chi(\zeta) = 
\overline{\chi} \left[ 1 - 
\sin \left( 2 \pi \sqrt{2} \frac{\zeta}{\lambda_\mathrm{u}}
\right ) \right ] .
\label{chizeta}
\end{equation}
The coordinate $\zeta$ measures the crystal thickness along 
the the direction of its growth. The beam has to be directed 
along the bent $(01\bar{1})$ plane at a small angle to the axis
$[011]$. In this case the beam makes the angle of approximately 
$45^{\circ}$ with the direction of the crystal growth and, therefore,  
the bending period of the channel is $\lambda_\mathrm{u}$.

The parameter $\overline{\chi}$ in (\ref{chizeta}) is the
average atomic fraction of germanium in the crystal.
For desired parameters of the channel bending $\lambda_\mathrm{u}$
and $a_\mathrm{u}$,
it can be found from the following formula \cite{Kostyuk2013}
\begin{equation}
\overline{\chi} = 170
\frac{a_\mathrm{u}}{\lambda_\mathrm{u}} \, .
\label{chibar0num}
\end{equation}
The values of $\overline{\chi}$ listed in table \ref{table}
correspond to the undulator parameters that were used in the simulations
presented in figures \ref{e-_spectra} and \ref{e+_spectra}.
It has been shown in \cite{Kostyuk2013} that strained layer 
crystals with such parameters are stable 
against misfit dislocations.
\begin{table}[tb]
\caption{The values of the  average atomic fraction of germanium
$\overline{\chi}$
that correspond to the parameters of the crystalline undulators  
that have been 
used in the simulations presented in figures
\ref{e-_spectra} and \ref{e+_spectra}. \label{table}}
\begin{tabular}{||c||c|c|c||}
\hline \hline
$a_\mathrm{u}$ &
\multicolumn{3}{c||}{$\lambda_\mathrm{u}$ (nm)} \\
\cline{2-4} 
(\AA) & $200$ & $400$ & $600$ \\
\hline \hline
\ \ 0.4 \ \ &\ \ 0.034 \ \ &\ \ 0.017 \ \ &\ \ 0.011 \ \ \\ 
\hline
\ \ 0.6 \ \ &\ \ 0.051 \ \ &\ \ 0.026 \ \ &\ \ 0.017 \ \ \\ 
\hline \hline
\end{tabular}
\end{table}

Hence, the crystalline undulator that violates conditions 
(\ref{large_a}) and (\ref{stable_ch}) can be created.
Moreover, it is predicted to be far superior 
to LALP CU.

Due to small $\lambda_\mathrm{u}$, the condition 
(\ref{Nu_Ld}) can be satisfied for SASP CU even 
if it is used with a moderate energy electron beam. This is 
a big practical advantage since positron sources are
much more expensive than electron ones, and electron 
beams are usually of higher quality.

In contrast to the LALP CU, the undulator radiation 
of SASP CU is much harder than the channeling 
radiation (\ref{large_omegau}). This a very 
important feature.
A soft photon background is usually easier
to get rid of. It is possible to make a detector 
sensitive to hard photons, but screen it from (or 
make insensitive to) soft photons. But it is more difficult 
 to do the opposite. This makes 
SASP CU substantially more suitable for many potential
applications in comparison to LALP CU.

Finally, due to much the smaller bending period of SASP CU,
it can produce by about 2 orders of magnitude harder 
photons when used with a beam of the same energy as
LALP CU. Or, to put it differently,  SASP CU will
require a much less expensive accelerator than would be needed
for production 
of photons of the same energy with LALP CU.
For the same reason, SASP CU may have an exciting technological
potential even in the domain of photon energies around $100$ keV
where synchrotron radiation sources do exist but are unique 
and very expensive.
\begin{acknowledgments}
The author is grateful to X.~Artru, H.~Backe, V.~Baryshevsky, S.~Dabagov, 
M.~Garattini, V.~Guidi, A.~Mazzolari, M.~Motapothula,
A.~Shchagin, N.~Shul'ga, 
V.~Tikhomirov, and W.~Wagner for their interest in the work, 
for encouraging discussions and critical comments.
At its initial stage, the work was supported by
Deutsche Forschungsgemeinschaft (DFG).
Numerical simulations were done at the 
Center for Scientific Computing of 
the J.W.~Goethe University, Frankfurt.
\end{acknowledgments}

\end{document}